\documentclass[conference]{IEEEtran}

\makeatletter

\def\ps@IEEEtitlepagestyle{%
  \def\@oddfoot{\mycopyrightnotice}%
  \def\@evenfoot{}%
}
\def\mycopyrightnotice{%
  {\footnotesize XXX-X-XXXX-XXXX-X/XX/\$XX.00~\copyright~20XX IEEE\hfill}% <--- Change here
  \gdef\mycopyrightnotice{}
}

\usepackage{blindtext}
\usepackage{eso-pic}
\IEEEoverridecommandlockouts
\usepackage{cite}
\usepackage{amsmath,amssymb,amsfonts}
\usepackage{algorithmic}
\usepackage{graphicx}
\usepackage{textcomp}
\usepackage{xcolor}
\def\BibTeX{{\rm B\kern-.05em{\sc i\kern-.025em b}\kern-.08em
    T\kern-.1667em\lower.7ex\hbox{E}\kern-.125emX}}
    
\usepackage{eso-pic}
\newcommand\AtPageUpperMyright[1]{\AtPageUpperLeft{%
 \put(\LenToUnit{0.17\paperwidth},\LenToUnit{-2cm}){%
     \parbox{0.9\textwidth}{\raggedleft\fontsize{8}{11}\selectfont #1}}%
 }}%
\newcommand{\conf}[1]{%
\AddToShipoutPictureBG*{%
\AtPageUpperMyright{#1}
}
}

\begin{document}
\title{\vspace*{1cm}A Comprehensive Evaluation of the Impact of ATM QoS Mechanisms on Network Performance for Multimedia and Data Applications
\\
{\footnotesize \textsuperscript{}}

}

\author{\IEEEauthorblockN{\textsuperscript{} Mahdi Manavi}
\IEEEauthorblockA{\textit{Department of Computer Science} \\
\textit{University of Houston}\\
Houston, USA \\
Mmanavi@uh.edu}

}

\maketitle
\conf{\textit{ 4. Interdisciplinary Conference on Electrics and Computer (INTCEC 2024) \\ 
11-13 June 2024, Chicago-USA}}

\begin{abstract}
The Asynchronous Transfer Mode (ATM) network is crucial due to its ability to efficiently transmit data, provide reliable connections, and support various service classes with specific Quality of Service (QoS) requirements. In this paper, we utilize the OPNET network simulation software to model an ATM network and analyze the impact of QoS classification on network performance. We investigate the effects of Constant Bit Rate (CBR), Variable Bit Rate (VBR), Available Bit Rate (ABR) and Unspecified Bit Rate (UBR) models on various network traffic types such as voice, video and data. For voice traffic, we examine key QoS parameters including Jitter, Packet Delay Variation and End-to-End Delay. For video traffic, we evaluate Packet Delay Variation and End-to-End Delay. Additionally, we analyze Download Response Time for data traffic to assess the influence of QoS on the ATM network.
Our results demonstrate that CBR and VBR are preferred for real-time traffic like voice and video, providing low delay and jitter. The simulation approach enables us to test various configurations and gain insights not possible in hardware tests. Our findings can help network operators determine the optimal QoS settings and tradeoffs when deploying ATM for modern multi-service networks.

\end{abstract}


\section{Introduction}
The capacity of network components to provide a certain service with a given level of assurance, improving performance and ensuring dependable data delivery, is known as QoS \cite{nnew1}.
QoS is critical for efficient and reliable data transmission in ATM networks. ATM networks function at the OSI model's data connection layer and are a type of high-speed networking technology \cite{1new}. QoS gives companies the ability to prioritize high-performance applications above other types of traffic on the network \cite{2new}.

ATM networks, characterized by fixed-length cells, support various service classes with specific QoS requirements to accommodate diverse traffic types. The importance of QoS lies in its ability to support real-time applications like voice and video communication, which demand consistent data rates and minimal delays \cite{1}.

ATM QoS is governed by several parameters that specify performance requirements for different traffic classes. These include user-defined attributes like CBR, VBR, ABR, and UBR, as well as network performance metrics such as Cell Loss Ratio (CLR), Cell Transfer Delay (CTD), Cell Delay Variation (CDV), and Cell Error Ratio (CER) \cite{2}.
However, ATM networks face challenges like packet loss, delay, and jitter due to packet switching, which are especially problematic for real-time and interactive traffic. To mitigate these issues, ATM employs traffic shaping to regulate incoming flows, traffic policing to monitor QoS compliance, and traffic engineering to optimize resource allocation for improved performance \cite{3}.

Proper QoS implementation is essential for ATM networks to prioritize critical traffic and maximize resource utilization. It ensures predictable and measurable service quality, crucial for delay-sensitive applications. Compliance with industry standards, particularly those set by the International Telecommunication Union (ITU), is vital for effective QoS implementation \cite{4}.
Despite its importance, the performance implications of using different ATM QoS service classes for mixed traffic scenarios are not well understood. There is a need for quantitative insights on the impact of QoS configurations on metrics like delay, jitter, and throughput for voice, video, and data traffic.
In this work, we leverage an OPNET simulation model of an ATM network to evaluate the performance of voice, video, and data traffic under different QoS classes. We quantify metrics including end-to-end delay, jitter, and response time to determine the optimal service class for each traffic type. The goal is to provide recommendations for selecting QoS parameters to improve real-time application performance while maintaining fairness and throughput for data traffic.

The key contributions of this paper are:\\
- An OPNET simulation model of an ATM network with configurable QoS classes\\
- Performance analysis of CBR, VBR, ABR, and UBR service classes\\
- Evaluation of delay, jitter, and response time for voice, video, and data traffic\\
- Guidelines for selecting the optimal ATM QoS class for different traffic types\\
The rest of the paper is organized as follows. Section 2 provides background on ATM QoS classes. Section 3 presents the simulation setup and methodology. Section 4 analyzes the results and performance impact. Section 5 concludes with key insights and recommendations.

\section{Related Work}
Several studies have investigated the performance implications of different QoS mechanisms in ATM networks through analytical models, simulations, and practical implementations. However, few have comprehensively analyzed the impact of ATM service classes on mixed voice, video, and data traffic using simulations.
Saltouros et al.\cite{5} proposed the use of a reinforcement learning algorithm to optimize routing in ATM networks based on the PNNI standard. Their approach aimed to maximize network throughput while guaranteeing QoS requirements for each connection. Considering the scalability challenges of large hierarchical ATM networks, the reinforcement learning algorithm efficiently allocated resources and optimized routing to meet QoS constraints. By intelligently adapting to network conditions through learning, their method could achieve high throughput performance while satisfying diverse QoS needs across connections in these complex networks. This work demonstrated the potential of applying machine learning techniques like reinforcement learning to jointly optimize throughput and QoS provisioning in ATM environments.
kumar et al. \cite{6} conducted a comprehensive study evaluating and comparing QoS support capabilities of different service classes in ATM networks. Their work analyzed the performance of ABR, CBR, and VBR schemes in terms of critical QoS metrics like transit delay and total end-to-end delay experienced by network traffic. By quantifying and contrasting the delay characteristics of these service classes, the researchers provided valuable insights into the QoS provisioning strengths and limitations of each scheme. This comparative analysis enabled a thorough understanding of how ABR, CBR, and VBR mechanisms facilitate QoS support in ATM environments, guiding the selection of appropriate service classes to optimize delay performance for different application requirements. The study's findings offer a solid foundation for configuring and tuning ATM networks to meet diverse QoS needs effectively.
Selvakumar et al. \cite{7} presented a protocol design aimed at enhancing QoS reliability and optimizing congestion probability routing in ATM networks. Their approach leveraged call admission control mechanisms to achieve QoS objectives such as maximizing network resource utilization while minimizing costs. The protocol's design principles focused on providing strong guarantees on critical QoS parameters like cell delay and loss probability. To validate their work, the authors implemented and integrated the proposed protocol with the OPNET simulation platform. This practical implementation facilitated comprehensive testing and evaluation of the protocol's performance in improving QoS assurance and congestion management in ATM network environments. By combining theoretical design principles with simulation-based implementation, the study offered insights into realizing QoS-aware routing and admission control strategies that can deliver reliable performance guarantees while optimizing resource usage in ATM infrastructures.

In contrast to these narrower studies, our work provides a comprehensive simulation-based evaluation of all major ATM service classes (CBR, VBR, ABR, UBR) and their impact on end-to-end delay, jitter, and response time for mixed voice, video, and data traffic scenarios. This holistic approach enables quantitative insights into selecting optimal QoS configurations for diverse application requirements in modern multi-service ATM networks.

\section{Implementation}
In the design of this network, multiple ATM switching components have been employed, utilizing 6 ATM8\_crossConn\_Adv cross-connect switches and a number of ATM\_client\_uni\_adv client switches. These components are organized into logical subnet sections for better network segmentation and management.

The network architecture incorporates dedicated stations for efficient voice and video transmission over the infrastructure. Additionally, separate data transmission stations have been configured to relay information through a centralized server, enabling seamless data exchange across the network.

The types of data traffic supported within this network encompass:\\
1. Email: Electronic mail communication.\\
2. File Transfer Protocol (FTP): Facilitating file transfers between clients and servers.\\
3. Voice: Real-time voice communication, such as VoIP calls.\\
4. Images/Video: Transmission of visual media, including still images and video streams.

To ensure proper QoS and prioritization, specific traffic classes have been assigned. Email and FTP data are classified as high-load traffic, receiving higher priority for reliable and efficient delivery. Voice traffic is set to PCM quality and speech settings, optimizing for real-time voice transmission. Video streams are configured with low-resolution settings to balance bandwidth requirements.

The network components are interconnected using DS1 (Digital Signal 1) data rate links, providing sufficient bandwidth for the various traffic types and enabling seamless communication between network devices.

The simulation duration for testing and evaluating the network performance is set to 220 seconds, allowing for comprehensive analysis of network behavior under various traffic loads and conditions.

\begin{figure}[htbp]
\centerline{\includegraphics[height=4.5cm, width=\linewidth]{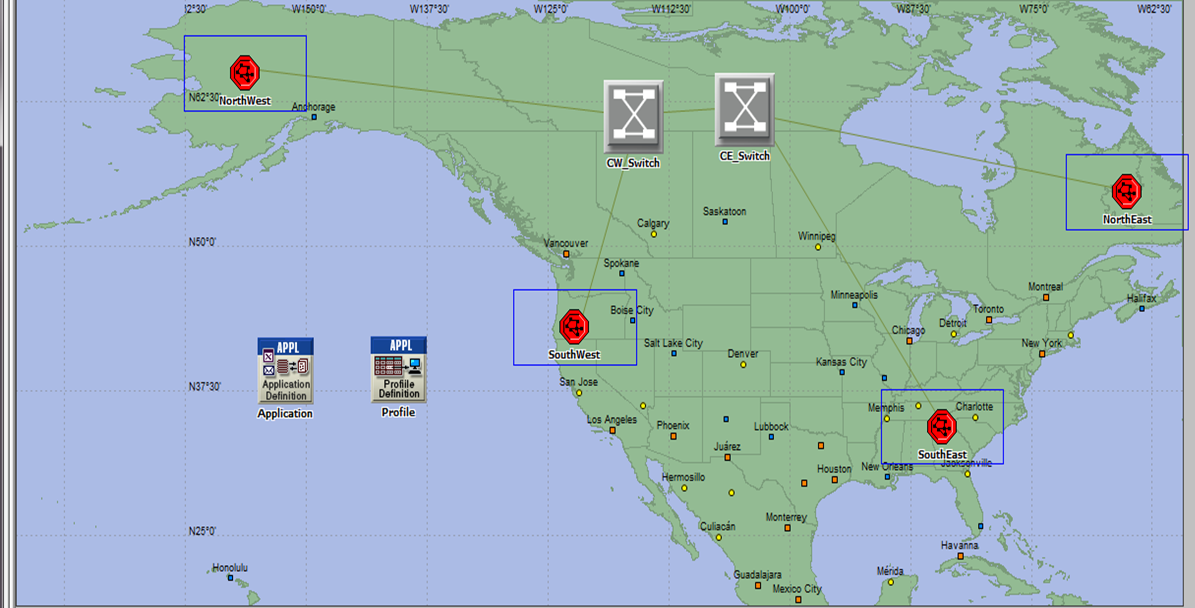}}
\caption{The network topology.}
\label{1}
\end{figure}

Figure \ref{1} shows that the base design of the network follows the structure outlined below, detailing the interconnections and logical layout of the components
This network architecture aims to provide a robust and efficient infrastructure for handling diverse data traffic, including email, file transfers, voice communication, and multimedia content, while ensuring appropriate quality of service and prioritization based on traffic types.
Furthermore, figure \ref{2} shows each of the SubNets:
\begin{figure}[htbp]
\centerline{\includegraphics[height=4.5cm, width=\linewidth]{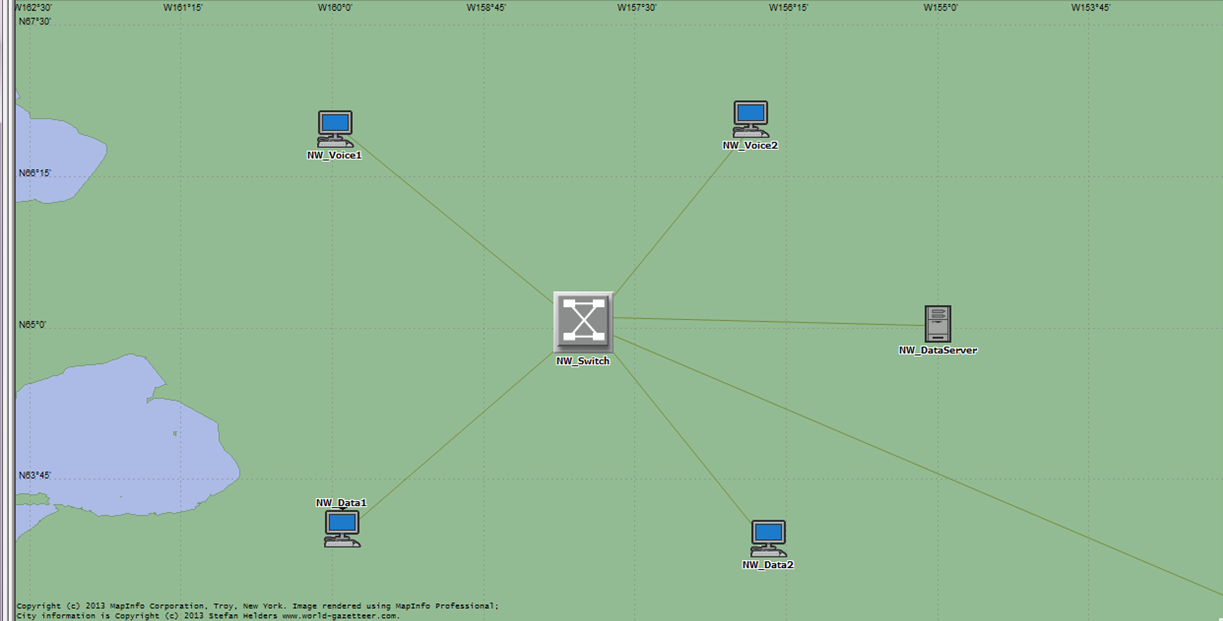}}
\caption{The SubNet topology.}
\label{2}
\end{figure}
By acting as logical divisions within the wider network architecture, the SubNets offer a methodical and well-organized way to segment networks. A distinct set of parts and features designed to manage particular kinds of data traffic or services make up each SubNet.

One SubNet, for instance, might be devoted to managing voice and video communications, complete with media servers, voice gateways, and the necessary switching equipment. Data transmission could be handled by a different SubNet that houses switches, routers, and servers designed for effective file transfers and data exchange.

The design provides various advantages by dividing the network into these logical SubNets, such as:\\
1. Improved network management: Network administrators can more easily monitor and maintain specific segments of the network, isolating issues and applying targeted configurations or policies.\\
2. Enhanced security: By implementing access controls and security policies unique to each subnet, network segmentation using SubNets reduces the possibility of extensive security breaches.\\
3. Traffic optimization: By using the proper bandwidth allotment, routing policies, and QoS settings, each SubNet may be configured to manage particular types of traffic. \\
4. Scalability: Without affecting the current infrastructure, more SubNets can be added to offer new services or meet higher traffic loads as network demands rise. 

A strong and effective communication infrastructure is facilitated by the SubNets' modular design, which offers flexibility, maintainability, and fine-grained control over many network elements.
In this simulation, the data stations receive their services from servers, and the voice stations are interconnected with each other. For all stations, the ATM Application Parameter is set based on which QoS class we use.
\subsection{Scenarios}
In these scenarios, different ATM service categories are employed to handle various types of network traffic. Each service category is designed to provide specific QoS characteristics, tailored to the requirements of the traffic being transmitted.
\subsubsection{Scenario A CBR}
In this scenario, all network traffic, including files, emails, voice, and images, is transmitted and received using the CBR service category. CBR is suitable for real-time, delay-sensitive applications that require a fixed amount of bandwidth to be continuously available. This service category guarantees a constant bit rate, ensuring that time-critical data, such as voice and video, can be delivered with minimal delay and jitter.
\subsubsection{Scenario B UBR}
In this scenario, data transmission occurs using the UBR service category. UBR is a best-effort service without any guaranteed bandwidth or delivery. It is typically used for non-critical data transfers, such as bulk file transfers or email transmissions, where timeliness is not a primary concern. UBR traffic is transmitted when bandwidth is available, after higher-priority traffic has been accommodated.
\subsubsection{Scenario C VBR}
The VBR service category is employed for network data transmission in this scenario. VBR is designed for bursty traffic with varying bandwidth requirements, such as multimedia streams or video conferencing. It provides a guaranteed bandwidth during bursts of data transmission while allowing for statistical multiplexing during periods of lower activity. VBR strikes a balance between the stringent requirements of CBR and the best-effort nature of UBR.
\subsubsection{Scenario D  ABR}
In this scenario, data is transmitted using the ABR service category. ABR is a feedback-based flow control mechanism that dynamically adjusts the transmission rate based on the available network resources. It aims to minimize cell loss and maximize utilization by allowing sources to share the available bandwidth fairly. ABR is suitable for data applications that can adapt to varying bandwidth availability, such as file transfers or email transmissions.

By examining these different scenarios, the simulation aims to evaluate the performance and behavior of the network under various traffic conditions and service categories. Each scenario will provide insights into how the network handles specific types of traffic, allowing for optimization, capacity planning, and the identification of potential bottlenecks or performance issues.
The results of these scenarios can help network administrators and designers make informed decisions regarding the appropriate service categories to use for different types of traffic, ensuring efficient resource utilization, meeting quality of service requirements, and delivering an optimal user experience across the network infrastructure.

\section{Results}
In this section, we will showcase the results and analysis derived from the simulations performed on the proposed network architecture. These simulations were designed to evaluate the network's performance and behavior under various scenarios, each employing different service categories and traffic configurations.
\subsection{Video}
In this section, we will delve into the comparative analysis of various video parameters across the four simulated scenarios, shedding light on the network's performance characteristics under different service class implementations.
\subsubsection{Packet Delay Variation}
Packet Delay Variation, also known as Jitter, is a crucial metric that measures the variability in the arrival times of packets within a data stream. This parameter is particularly relevant for time-sensitive applications, such as video transmission, as excessive jitter can lead to degraded quality, frozen frames, or disruptions in the viewing experience.
In this analysis, we investigate the Packet Delay Variation observed for each service class across the four scenarios. The corresponding graph, presented below, visually represents the comparative performance of the service classes in terms of jitter.

\begin{figure}[htbp]
\centerline{\includegraphics[height=4.5cm, width=\linewidth]{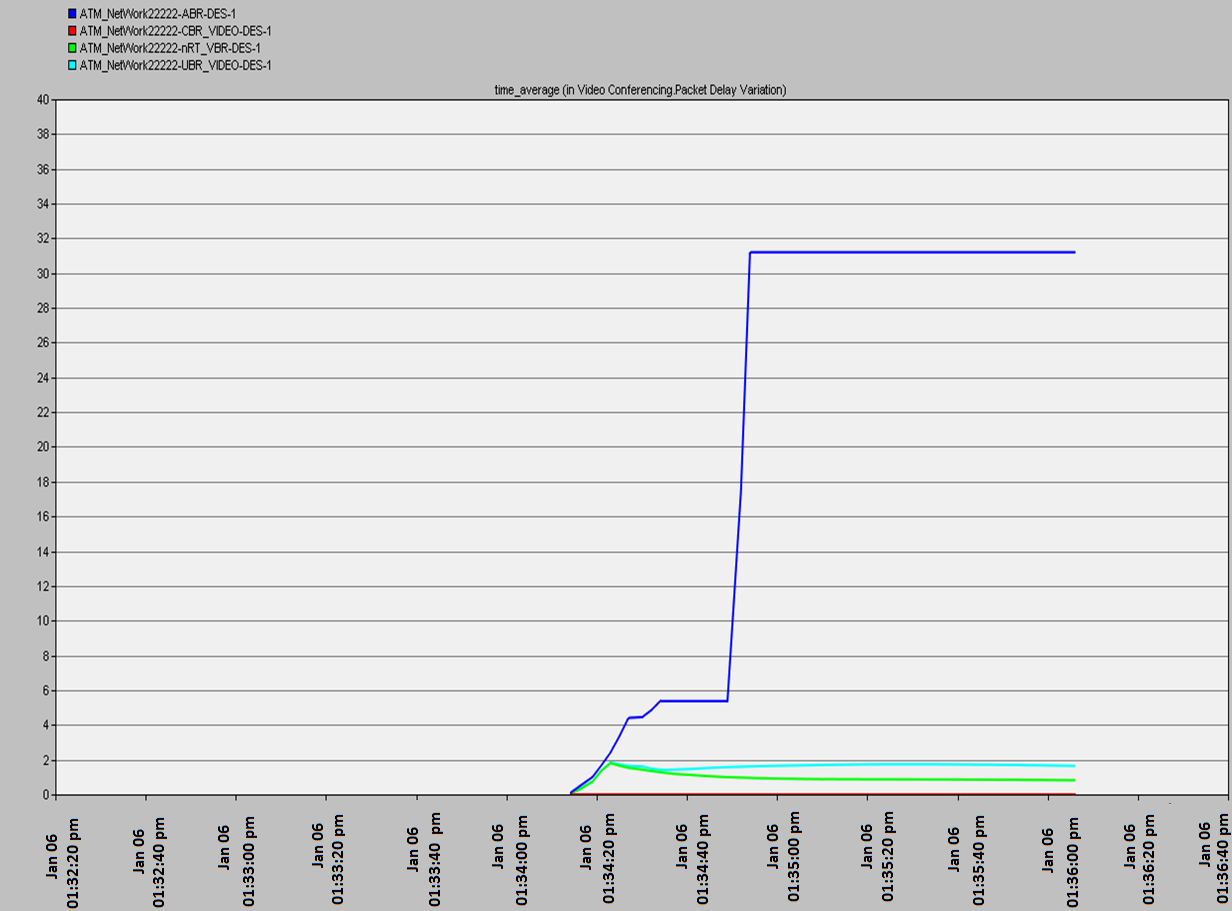}}
\caption{Packet delay variation for 4 services.}
\label{3}
\end{figure}

Fig \ref{3} clearly illustrates the superiority of the CBR service class in terms of minimizing Packet Delay Variation, making it the preferred choice for high-quality video transmission over the ATM network infrastructure. VBR also demonstrates relatively low jitter levels, making it a viable option for certain video applications with varying bandwidth requirements. ABR and UBR, while offering different levels of resource utilization and adaptability, may not be suitable for delay-sensitive video applications due to their higher observed Packet Delay Variation.
This analysis provides valuable insights into the selection of appropriate service classes for video transmission, enabling network administrators and designers to make informed decisions based on the specific requirements and constraints of their video applications and overall network infrastructure.

\subsubsection{Packet End-to-End Delay}
\begin{figure}[htbp]
\centerline{\includegraphics[height=4.5cm, width=\linewidth]{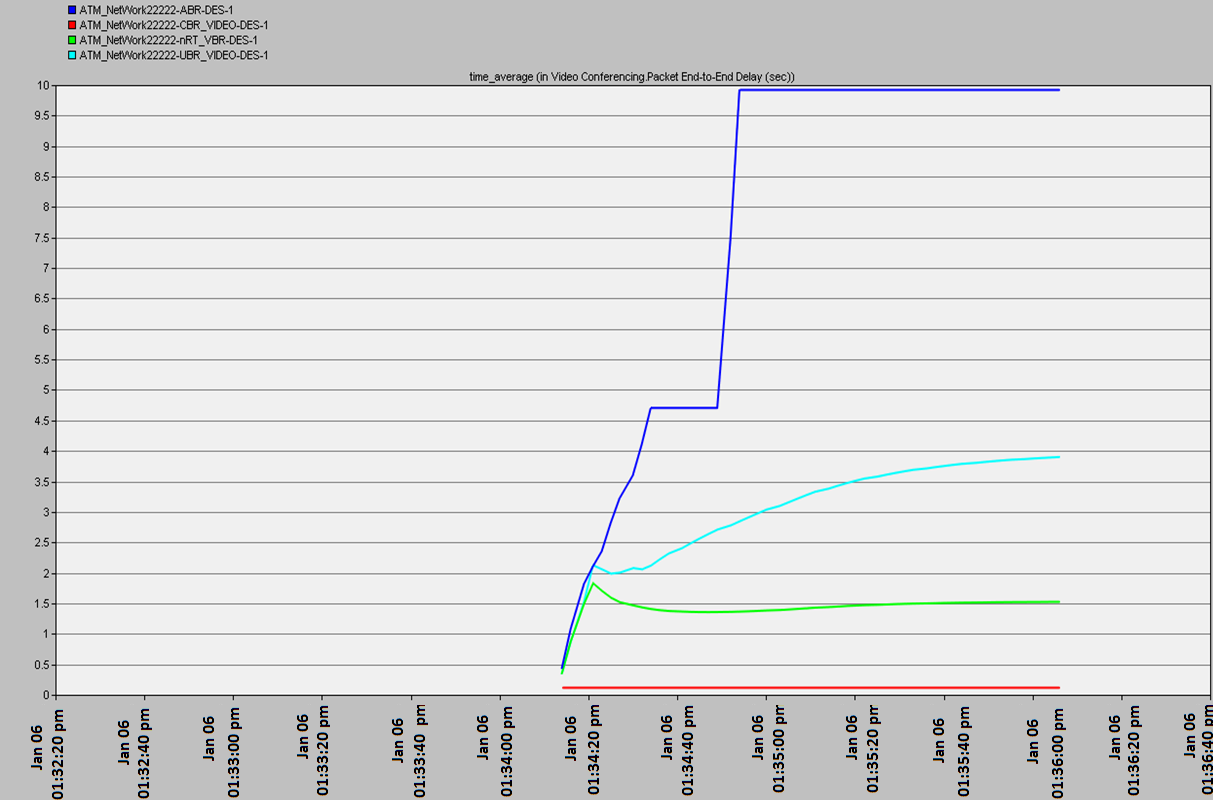}}
\caption{Packet End-to-End Delay for 4 services.}
\label{4}
\end{figure}

Fig \ref{4} demonstrates the superiority of the CBR service class in minimizing Packet End-to-End Delay, making it the preferred choice for applications that require timely packet delivery, such as real-time video streaming or VoIP communication. VBR also exhibits relatively low delay levels, making it a viable option for applications with varying bandwidth requirements. ABR and UBR, while offering different levels of resource utilization and adaptability, may not be suitable for delay-sensitive applications due to their higher observed end-to-end delays.
Based on fig 3 and Fig 4, it can be concluded that the best class in terms of the two considered variables for implementation is the CBR class.

In table 1, we have provided various parameters for video in numerical form for each chart in a table.
\begin{table}[htbp]
    \centering
    \caption{Numerical analysis of video transfer services.}
    \label{tab:costs}
    \begin{tabular}{|c|c|c|}
        \hline
       
        \textbf{Services}& \textbf{Packet Delay Variation(s) } & \textbf{End-End Delay(s)} \\
        \hline
        CBR&	[0.00000002,0.00000011]	& [0.11,0.11]
        \\
         \hline
        
       UBR&	[0.05,1.82] &	[0.47,3.9] \\
          \hline 
        VBR	& [0.08,1.8]	& [0.38,1.8]
        \\
        \hline 
       ABR	& [0,28] &	[0.8,10]\\
        \hline
    \end{tabular}
\end{table}

\subsection{Voice}
In this section, we will examine various parameters for voice data
\subsubsection{Jitter}
\begin{figure}[htbp]
\centerline{\includegraphics[height=4.5cm, width=\linewidth]{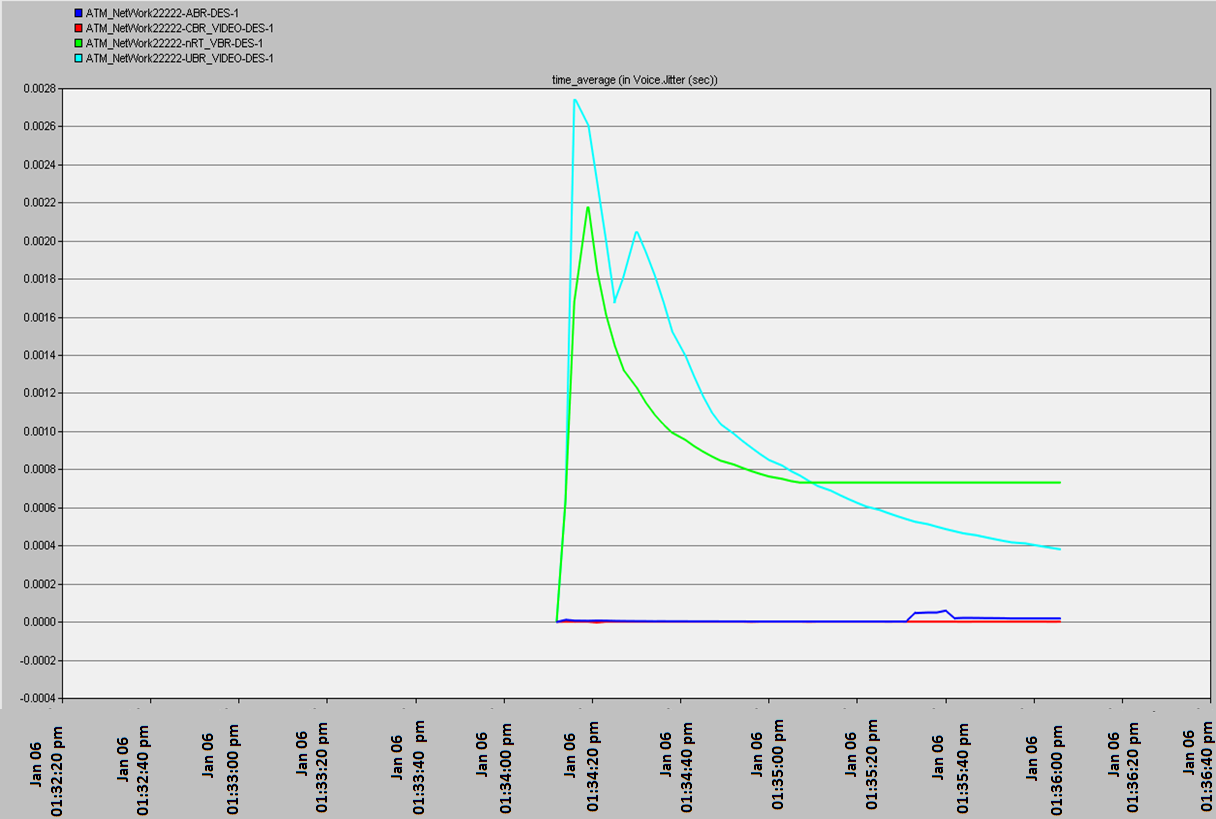}}
\caption{Jitter for 4 services.}
\label{5}
\end{figure}

In the initial section where we analyze audio data, we focus on the Jitter parameter for this type of data. Below, we compare four services in figure \ref{5}.

Fig \ref{5} amply demonstrates the CBR service class's superiority in reducing Jitter, which makes it the recommended option for applications like VoIP and real-time video streaming that are extremely sensitive to changes in packet delivery timings. Additionally, VBR has comparatively low jitter levels, which makes it a good choice for applications that require changing bandwidth but can tolerate moderate jitter. Because of their greater reported jitter levels, ABR and UBR may not be appropriate for real-time or delay-sensitive applications, even if they offer varying degrees of flexibility and resource usage.
Through this study, network administrators and designers can better understand how to choose the right service classes depending on the jitter requirements of various applications, optimizing the network infrastructure and guaranteeing a top-notch user experience for delay-sensitive applications.
\subsubsection{Packet Delay Variation}
\begin{figure}[htbp]
\centerline{\includegraphics[height=4.5cm, width=\linewidth]{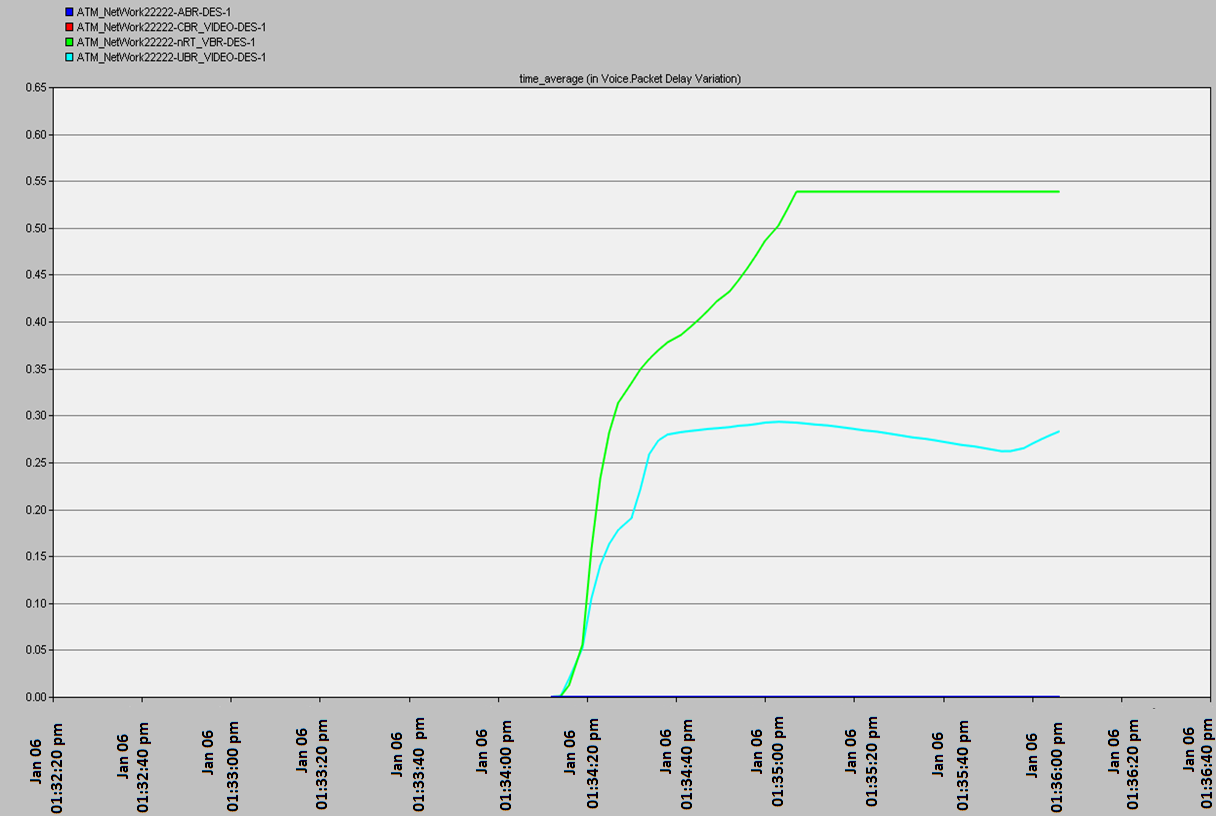}}
\caption{Packet delay variation for 4 services.}
\label{6}
\end{figure}

In Figure \ref{6}, we illustrate the packet delay variation for each service.

Figure \ref{6} demonstrates the superiority of the CBR service class in minimizing Packet Delay Variation for video traffic, making it the preferred choice for high-quality video streaming applications that demand smooth and uninterrupted playback. VBR also exhibits relatively low jitter levels, making it a viable option for video applications with varying bandwidth requirements but moderate jitter tolerance. ABR and UBR, while offering different levels of resource utilization and adaptability, may not be suitable for delay-sensitive or high-quality video applications due to their higher observed jitter levels, which can degrade the overall video viewing experience.
This analysis provides valuable insights into the selection of appropriate service classes specifically for video applications, enabling network administrators and designers to optimize the network infrastructure and ensure a high-quality video streaming experience for users.
\subsubsection{End-to-End Delay}
In this section, we focus on examining this parameter to analyze the performance of the classes.
\begin{figure}[htbp]
\centerline{\includegraphics[height=4.5cm, width=\linewidth]{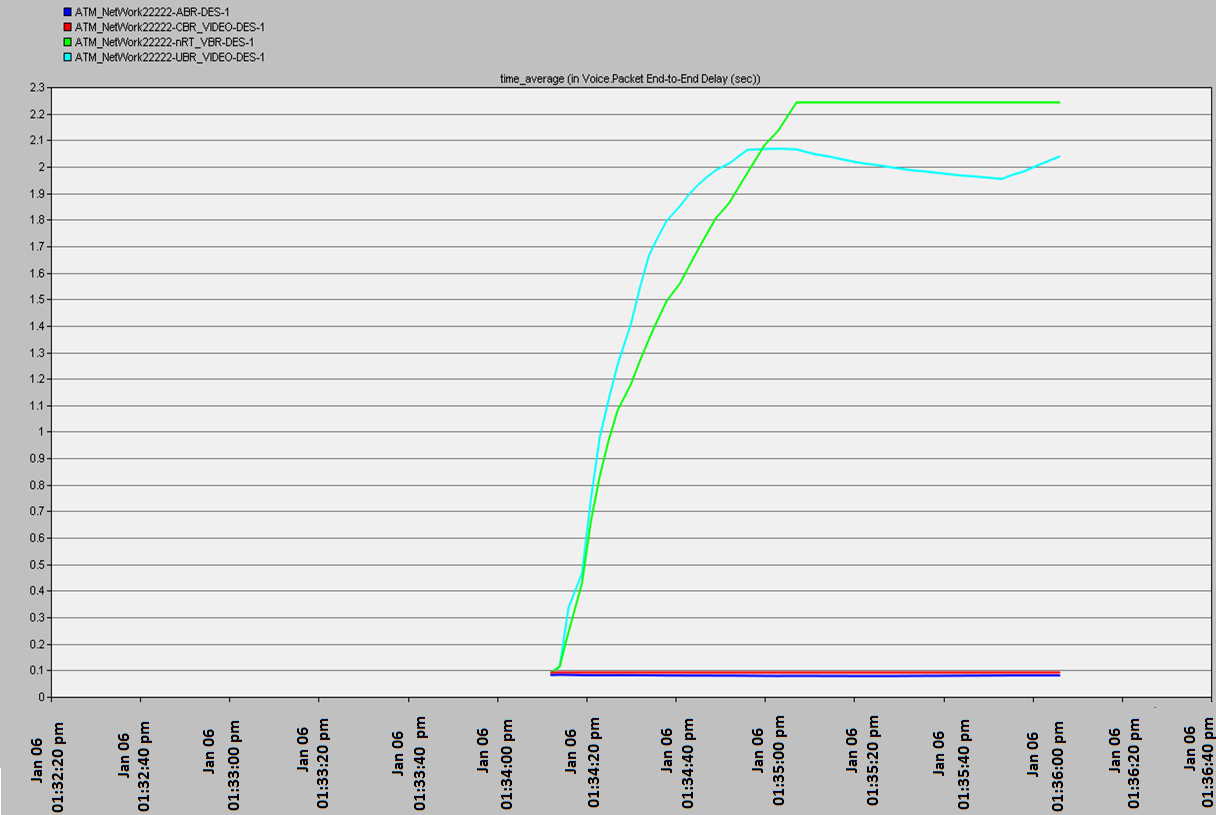}}
\caption{End-to-End Delay for 4 services.}
\label{7}
\end{figure}

figure \ref{7} shows that the ABR service class unexpectedly outperforms other classes, including CBR, in terms of minimizing End-to-End Delay for voice traffic in this specific scenario. CBR, being designed for real-time applications, still exhibits relatively low delays, making it a viable option for voice communication. VBR and UBR may not be optimal choices for delay-sensitive voice applications due to their higher observed end-to-end delays.
In table 2, we present the numerical values of each service in voice transfer to provide a clearer understanding.
\begin{table}[htbp]
    \centering
    \caption{Numerical analysis of voice transfer services.}
    \label{tab:costs}
    \begin{tabular}{|c|c|c|c|}
        \hline
       
        \textbf{Services}& \textbf{Jitter(s) } & \textbf{Packet Delay Variation(s)} &\textbf{END-END Delay(s)}\\
        \hline
       CBR	& 0 &	[0.0000009,0.0000018] &	[0.91,0.91]
        \\
         \hline
        
       UBR	& [0, 0.0027] &	[0,0.0015]	& [0.1,2.1] \\
          \hline 
        VBR	& [0,0.0021] & [0,0.54]	& [0.1,2.25]
        \\
        \hline 
      ABR	& [0,0.000005] &	[0,0.0000016] &	[0.078,0.081]\\
        \hline
    \end{tabular}
\end{table}

\subsection{Email}
In this section, we will analyze the Download Response Time metric across the four service classes employed in our simulations.

The download response time is a crucial performance indicator that measures the time it takes for a client to receive a complete download from a server. This metric is particularly relevant for file transfer applications, as it directly impacts the user experience and perceived responsiveness of the network.
\begin{figure}[htbp]
\centerline{\includegraphics[height=4.5cm, width=\linewidth]{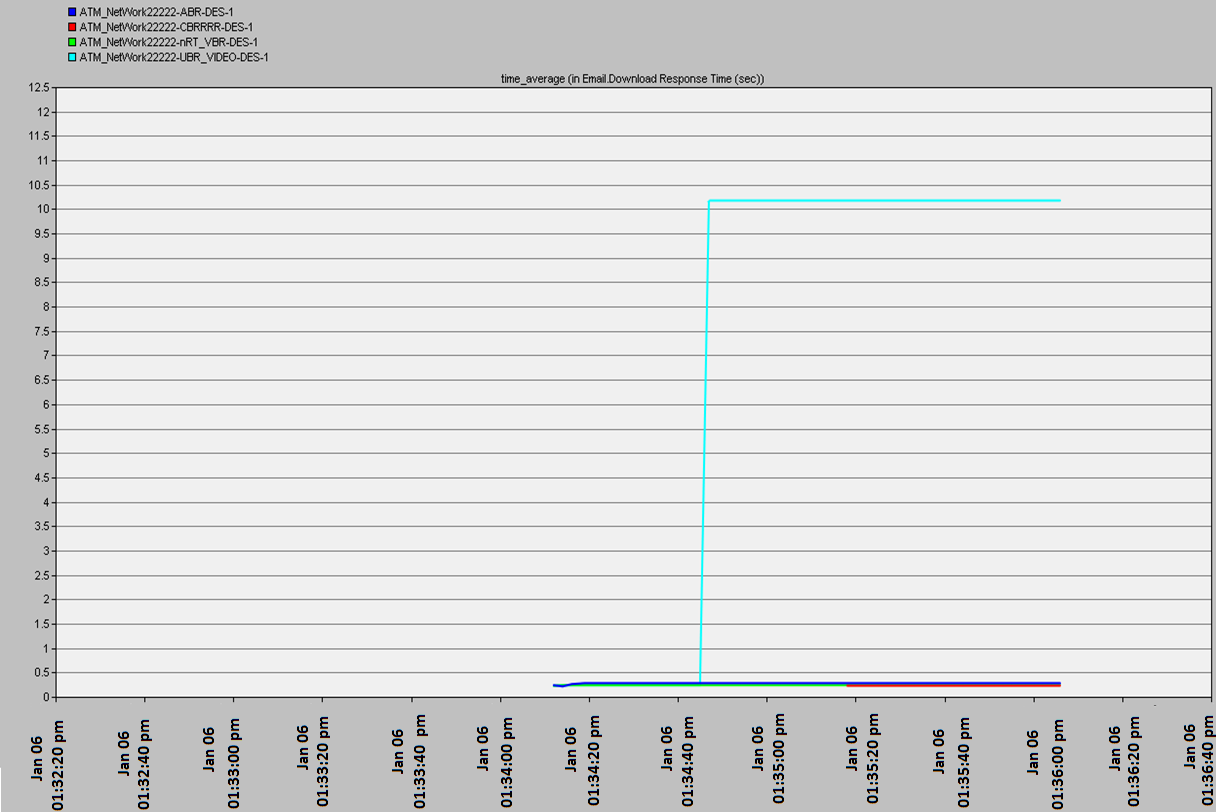}}
\caption{Download response time for 4 services.}
\label{8}
\end{figure}

Figure \ref{8} demonstrates that the CBR service class performs significantly better than the other classes and has the lowest download response time for email traffic. The ABR and Unspecified Bit Rate UBR classes show greater Download Response Times, which makes them less appropriate for effective email download operations. The VBR class performs the second-best. Overall, the graph emphasizes how crucial it is to choose the right service class depending on the needs of the application because it can have a big impact on network speed and user experience when it comes to file transfers via email.

Table 3 shows the numerical values of each service.
\begin{table}[htbp]
    \centering
    \caption{Numerical analysis of Email transfer services.}
    \label{tab:costs}
    \begin{tabular}{|c|c|}
        \hline
       
        \textbf{Services}& \textbf{Download response time(s)} \\
        \hline
        	CBR & [0.23, 0.23]
        \\
         \hline
        
       UBR&	[0.23,0.24] \\
          \hline 
        VBR	& [0.22,0.27]
        \\
        \hline 
       ABR	& [0.1,10.1]\\
        \hline
    \end{tabular}
\end{table}

\subsection{File Transfer Protocol (FTP)}
In this section, we will examine the performance of file data transfers, where workstations receive data and files from servers, focusing specifically on the Download Response Time parameter.

\begin{figure}[htbp]
\centerline{\includegraphics[height=4.5cm, width=\linewidth]{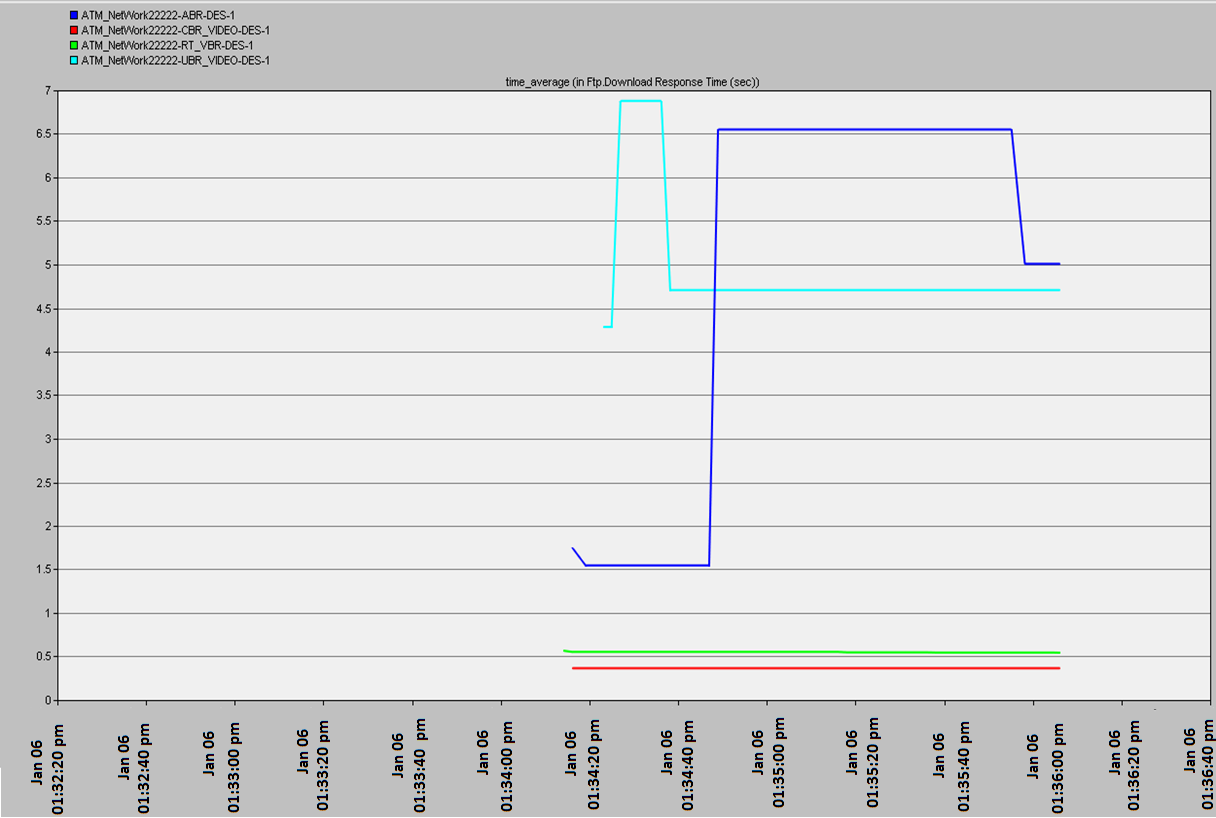}}
\caption{Download response time for 4 services.}
\label{9}
\end{figure}

Figure \ref{9} illustrates the behavior of the Download Response Time across different service classes for file transfer operations within the network infrastructure.

Efficient file transfer is a critical requirement for many applications and workflows within modern networks. The ability to quickly and reliably download files from servers to client workstations can significantly impact productivity, collaboration, and overall user experience.
The graph unequivocally showcases the preeminence of the CBR service class in delivering the shortest Download Response Time for FTP operations, rendering it the optimal choice for applications necessitating swift and dependable file transfers. While the VBR class also exhibits relatively low response times, making it a viable alternative for FTP applications with fluctuating bandwidth requirements, the ABR and UBR classes may not be well-suited for time-critical or large-scale file transfers over FTP due to their higher observed response times.

This analysis furnishes network administrators and designers with invaluable insights, enabling them to judiciously select the appropriate service class tailored to FTP and file transfer applications. Consequently, they can optimize the network infrastructure to deliver efficient and responsive file transfer services, thereby enhancing the overall user experience.

Table 4 shows the numerical values of each service.

\begin{table}[htbp]
    \centering
    \caption{Numerical analysis of FTP services.}
    \label{tab:costs}
    \begin{tabular}{|c|c|}
        \hline
       
        \textbf{Services}& \textbf{Download response time (s)} \\
        \hline
        	CBR & [0.36, 0.36]
        \\
         \hline
        
       UBR&	[0.52,0.52] \\
          \hline 
        VBR	& [1.51,6.51]
        \\
        \hline 
       ABR	& [4.30,6.90]\\
        \hline
    \end{tabular}
\end{table}

\section{Conclusion}

In this paper, we have presented a comprehensive simulation-based study evaluating the impact of ATM QoS classes on network performance for various traffic types, including voice, video, and data. Through extensive simulations using the OPNET network simulation software, we have quantified the performance characteristics of CBR, VBR, ABR, and UBR service classes across multiple QoS metrics.

Our results demonstrate that the CBR service class is the preferred choice for delay-sensitive and real-time applications, such as voice and video transmission, as it exhibits jitter and end-to-end delay. The VBR service class also performs well for applications with varying bandwidth requirements, offering relatively low jitter and delay levels. These findings highlight the suitability of CBR and VBR for time-critical applications that demand smooth and uninterrupted communication.

Furthermore, our analysis of the Download Response Time metric for data traffic, including email and FTP operations, reinforces the superiority of the CBR service class in delivering efficient and responsive file transfers. The VBR class also demonstrates relatively low response times, making it a viable alternative for applications with varying bandwidth requirements.

The simulation approach employed in this study enabled us to test various configurations and scenarios, providing quantitative results and analysis that would be challenging to obtain through hardware tests alone. Our findings can serve as a foundation for future research and development in the area of QoS management and traffic engineering in ATM networks.

In future work, we can work on integration with emerging technologies. Evaluating the interoperability and performance of ATM QoS mechanisms with emerging technologies, such as 5G networks, software-defined networking (SDN), and network function virtualization (NFV), would be crucial for enabling seamless integration and ensuring QoS support in future communication infrastructures.

\vspace{12pt}

\end{document}